\documentstyle[amssymb,aps]{revtex}

\begin{document}
\author{R.S. Freitas* and L. Ghivelder}
\address{Instituto de F\'{i}sica, Universidade Federal do Rio de Janeiro, C.P. 68528,%
\\
Rio de Janeiro, RJ 21945-970, Brazil}
\author{F. Damay}
\address{Materials Department, Imperial College, Prince Consort Road, London SW7 2BP,%
\\
UK}
\author{F. Dias}
\address{Instituto de F\'{i}sica, Universidade Federal do Rio Grande do Sul, C.P.\\
15051, Porto Alegre, RS 91501-970, Brazil}
\author{L.F. Cohen}
\address{Blackett Laboratory, Imperial College, Prince Consort Road, London SW7 2AZ,\\
UK}
\title{Magnetic relaxation phenomena and cluster glass properties of La$_{0.7-x}$Y$%
_{x}$Ca$_{0.3}$MnO$_{3}$ manganites}
\maketitle

\begin{abstract}
The dynamic magnetic properties of the distorted perovskite system La$%
_{0.7-x}$Y$_{x}$Ca$_{0.3}$MnO$_{3}$ (0 $\leq $ {\it x} $\leq $ 0.15) have
been investigated by ac-susceptibility and dc magnetization measurements,
including relaxation and aging studies. All investigated samples display a
metal-insulator transition. As yttrium is added in the compounds the overall
results show evidence for the gradual appearance of a cluster glass
behavior. For the {\it x} = 0.15 sample, magnetization measurements as a
function of time at various temperatures show that the magnetic relaxation
is maximum at a given temperature, well below the ferromagnetic transition.
This maximum coincides in temperature with a frequency dependent feature in
the imaginary part of the ac susceptibility, associated with a freezing
process. This is interpreted as due to ferromagnetic clusters, which grow
with decreasing temperature down to a temperature at which they freeze due
to severe intercluster frustration.

\end{abstract}

\section{Introduction}

Extensive study of the doped perovskite manganites La$_{1-y}$Ca$_{y}$MnO$%
_{3} $ revealed a very rich and complex phase diagram, which includes
colossal magnetoresistance (CMR) behavior near the transition temperature of
the metallic ferromagnetic compositions, 0.15%
\mbox{$<$}%
{\it y}%
\mbox{$<$}%
0.50.\cite{Coey} The ferromagnetic (FM) interaction appears as a consequence
of the manganese mixed valence state. Substitution of La$^{3+}$ by Ca$^{2+}$
in antiferromagnetic LaMnO$_{3}$ changes the Mn$^{3+}$/Mn$^{4+}$ ratio and
produces holes in the e$_{g}$ orbitals. The simultaneous para-ferromagnetic
and metal-insulator transitions are basically understood within the
framework of the double-exchange theory \cite{Zener}, which considers the
transfer of one e$_{g}$ electron between neighboring Mn$^{3+}$ and Mn$^{4+}$
ions through the path Mn-O-Mn.

For a given e$_{g}$ concentration the tolerance factor of the perovskite
structure, proportional to the average ionic size of the A site, was shown
to have a drastic influence on the physical properties of manganites.
Compounds of the type (La$_{0.7-x}$R$_{x}$)Ca$_{0.3}$MnO$_{3}$ (R = Pr, Y,
Dy, or Tb), where the Ca concentration is close to an optimum value in
relation to the FM interaction, are particularly suited to investigate this
effect: replacing La$^{3+}$ with smaller rare-earth ions decreases the
Mn-O-Mn bond angle, and thus affects the hopping of the e$_{g}$ electrons
and weakens the double-exchange interaction, while the antiferromagnetic
superexchange term is only marginally affected and therefore increase in
importance.\cite{Hongsuk} Consequently, both double exchange and
superexchange interactions will compete more strongly depending on the
structural distortion, giving rise to a magnetically disordered state. A
magnetic and electronic phase diagram of several manganite systems plotted
against the tolerance factor has been published on several investigations.%
\cite{DeTere,Terai} Lowering the value of the average ionic size of the A
site initially reduces the transition temperature T$_{C}$, and below a
critical value the low temperature phase changes to a spin-glass insulating
state. Preliminary studies\cite{Xu} have additionally shown that the FM
metallic phase of La$_{0.7-x}$Y$_{x}$Ca$_{0.3}$MnO$_{3}$ ({\it x} = 0.15)
display signatures of a magnetic cluster glass state.

The purpose of the present work is to investigate the dynamic properties of
this disordered phase by means of magnetic relaxation measurements. One
characteristic feature of glassy systems is that the magnetization of a
sample cooled in zero field to a certain temperature depends on how long it
is held in zero field before the magnetic field is applied. This is the so
called aging effect. Furthermore, if M({\it t}) is differentiated with
respect to ln{\it t}, the derivative S = $\partial $M/$\partial $(ln{\it t})
displays a maximum which shifts to higher {\it t} for longer values of the
wait time, {\it t}$_{w}$. These effects have been observed on a well
characterized cluster glass compound, the cobaltite La$_{0.5}$Sr$_{0.5}$CoO$%
_{3}$, where glassy behavior and intracluster ferromagnetism were shown to
coexist.\cite{Mukherjee,Nan} More recently, on Nd$_{0.7}$Sr$_{0.3}$MnO$_{3}$%
, it was shown that disorder and frustration may occur in the ferromagnetic
phase.\cite{NorbNdSr}

Within this context, this paper presents low temperature magnetic relaxation
measurements on La$_{0.7-x}$Y$_{x}$Ca$_{0.3}$MnO$_{3}$, with 0 $\leq $ {\it x%
} $\leq $ 0.15. As yttrium is added in the compound, the results evidence
for the first time the gradual appearance of glassy signatures within the
ferromagnetic metallic phase, prior to the existence of the insulating spin
glass phase which occurs with higher rare-earth doping at the La site.\cite
{DeTere,Terai} Relaxation measurements probe the out-of-equilibrium state of
the system, where the clear observation of a peak in the magnetic relaxation
as a function of temperature and the observation of aging effects further
characterize the cluster glass properties of this system.

\section{Experimental}

The investigated compounds are polycrystalline samples of La$_{0.7-x}$Y$_{x}$%
Ca$_{0.3}$MnO$_{3}$, with {\it x} = 0, 0.07, 0.10 and 0.15, prepared by
standard solid state reaction. X-ray analysis confirmed a single phase
orthorhombic perovskite structure. Magnetic measurements were made with an
extraction magnetometer (Quantum Design PPMS) when the applied field was
greater then 5 mT. Low field data, H = 0.3 mT, was taken with a SQUID
magnetometer (Quantum Design MPMS). Zero field cooled relaxation
measurements were performed on all samples at different temperatures between
10K and 250 K, and for applied fields varying from 5mT to 0.5T. The samples
were cooled from a reference temperature in the paramagnetic state with a
constant cooling rate, and kept at a target temperature for a certain wait
time {\it t}$_{w}$. Thereafter, a dc field was applied and the magnetization
was recorded vs. the elapsed time, {\it t}. The samples were additionally
characterized by transport measurements, made by a conventional ac
four-probe method. The results are shown on Fig. 1. All samples display a
metal-insulator transition, below which a metallic-like electrical
conductivity is observed. A similar behavior was previously reported for
these compounds.\cite{Obra}

\section{Results and Discussion}

To recall the characteristic magnetic behavior of La$_{0.7-x}$Y$_{x}$Ca$%
_{0.3}$MnO$_{3}$ compounds, Fig. 2 displays the zero field cooled (ZFC) and
field cooled (FC) temperature dependence of the dc magnetization of the
samples {\it x} = 0, 0.07, 0.10 and 0.15, measured at H =5 mT. The small
hump visible at low temperatures in most of the FC data is due to a problem
in the experimental apparatus. The results of Fig. 2 clearly show a standard
FM transition for the {\it x} = 0 sample at T$_{C}$ = 252 K, as observed on
numerous investigations.\cite{Coey} For the doped compounds ({\it x }$>$ 0)
the ZFC curves evolve continuously, with higher yttrium concentration, to a
cusp-like anomaly at a temperature T$_{a}$ just below the Curie-like
temperature T$_{C}$, defined as the maximum inflection of the FC data. The T$%
_{C}$ values obtained are 143, 114, and 89 K, for {\it x} = 0.07, 0.10, and
0.15 respectively. An important point to be noted here is that the FC
magnetization continues to increase strongly below the irreversibility
temperature, T$_{r}$, at which ZFC and FC curves merge, a typical feature of
various cluster glass systems.\cite{Koyano,Dusan} This effect is more
pronounced on the {\it x} = 0.15 sample. On the other hand, in canonical
spin glass systems the FC magnetization shows a nearly constant value below T%
$_{r}$. It worth noting that due to the fact that cluster glasses exhibit
finite range ferromagnetic ordering below T$_{C}$, this system may show some
features similar to those found in reentrant spin glass systems (RSG), which
undergo a PM-FM transition at T$_{C}$ and have a lower freezing temperature.
Nevertheless, it may be noted that in several RSG systems, such as NiMn\cite
{NiMn1}\cite{NiMn2}, AuFe\cite{AuFe}, and FeZr\cite{kaul}, the
irreversibility temperature T$_{r}$ occurs at temperatures well below T$_{C}$
(when the measuring field is higher then the coercive field), whereas in a
cluster glass the irreversibility arises just below T$_{C}$\cite{Mukherjee},
as observed in Fig. 2 for the yttrium doped samples.

On the other hand, magnetization curves measured as a function of field,
M(H), do show signatures of a RSG behavior. The low field portion of a
typical hysteresis loop measured after ZFC condition at 2.0 K, a temperature
well below T$_{r}$, is shown in the inset of Fig. 3. The curve displays an 
{\it S} shape in the virgin branch, with a positive curvature at low fields,
a typical characteristic of both canonical SG and RSG systems.\cite
{Mukherjee,Masayuki,senoussi} In the same way, the variation of the coercive
field H$_{C}$ with temperature, plotted in Fig. 3, is quite similar to those
usually observed in RSG systems, where H$_{C}$ increases rapidly as the
temperature is lowered below T$_{C}$. \cite{senoussi,campbell} Such
conflicting results evidence a more complex glassy behavior in these
compounds.

Susceptibility measurements are a very efficient way to evidence glassy
behavior. The temperature dependence of the ac susceptibility of La$_{0.55}$Y%
$_{0.15}$Ca$_{0.3}$MnO$_{3}$ is presented on Fig. 4. The variation with
temperature of the in-phase susceptibility $\chi $', shown on the inset of
the figure, is comparable to that of the dc M$_{ZFC}$, with an onset around
120K related to the paramagnetic to ferromagnetic transition, followed by a
maximum at T = 79 K, close to the M$_{ZFC}$ cusp temperature T$_{a}$ = 72 K.
This maximum is not frequency dependent, which suggests that the FM state
originates from intracluster ferromagnetism. Moreover, if the $\chi $'
maximum and T$_{a}$ are related, the cusp on the M$_{ZFC}$ curve is the dc
signature of these FM interactions, rather than a consequence of cluster
freezing. The out-of-phase susceptibility $\chi $'' also shows a maximum at
T = 87 K, which is frequency independent. It has been suggested\cite{Nan}
that the temperature at which this maximum occurs is related to the
irreversibility temperature, T$_{r}$ = 88 K, obtained on the dc
magnetization curves. Therefore the maximum in $\chi $'' at higher
temperatures is also related the intracluster FM interactions. However, an
additional feature clearly visible in $\chi $'' is the presence of a hump,
at T$_{f}$ $\thicksim $40K. This broad peak is frequency dependent and
shifts towards higher temperatures with increasing frequency, a
characteristic feature of the dynamics of spin glass systems. A similar
behavior has previously been reported in other manganite samples\cite{maig}.
This double peak structure in $\chi $'' is progressively less visible for
lower Y doping, and it is not present in the Y=0 sample.

In order to gain further information on the underlying nature of this
cluster glass system, we have measured the long time relaxation of the
magnetization, with a time scale greater then 10$^{4}$ s. Figure 5a displays
the normalized M({\it t})/M(0) data for all samples measured at a reduced
temperature T/T$_{C}$ = 0.2, with H = 5 mT (this probing field is within the
linear response regime). As Y is added in the compounds the data show a
gradual increase of M({\it t})/M(0), at any given time. Quantitatively,
after 10$^{4}$ seconds, the fractional change of the magnetization is 0.18,
0.93, 4.5, and 12\% in the samples with {\it x} = 0, 0.07, 0.10, and 0.15
respectively. Relaxation data at different temperatures for La$_{0.55}$Y$%
_{0.15}$Ca$_{0.3}$MnO$_{3}$ are shown in Fig 5b. The slope of M({\it t}%
)/M(0) increases below T$_{C}$, reaches a maximum at T/T$_{C}$ $\simeq $
0.4, and decreases for lower temperatures. This evolution of the
magnetization relaxation behavior with temperature is indicative of the
development of the magnetic clusters as temperature is lowered. It is
noteworthy that relaxation is observed not only below the freezing
temperature T$_{f}$, but also up to the FM transition temperature T$_{C}$,
which indicates that the system is not in a true ferromagnetic state. This
is reinforced with measurements of M(H) above and below T$_{C}$ (not shown),
where Arrot plots M$^{2}$ vs. H/M do not yield straight lines. Previous
observations in RSG system\cite{NordReen} also show that relaxation is
present above the freezing temperature of the spin glass state.

Amongst the various functional forms that have been proposed to describe
magnetization as a function of time, one of the most popular is a stretched
exponential of the form :

\begin{center}
\begin{equation}
M(t)=M_{0}-M_{r}\exp \left[ -\left( \frac{t}{\tau _{r}}\right) ^{1-n}\right]
\end{equation}
\end{center}

where M$_{0}$ relates to an intrinsic FM component, and M$_{r}$ to a glassy
component mainly contributing to the relaxation effects observed. The time
constant $\tau _{r}$ and the parameter {\it n} are related to the relaxation
rate of the spin-glass like phase. The values of {\it n} are scattered
between 0.48 and 0.6, in agreement with previous results.\cite{maig} The
variations of M$_{0}$ and M$_{r}$ with reduced temperature are shown on Fig.
6 for the samples with {\it x} = 0.10 and 0.15. As expected, M$_{0}$ and M$%
_{r}$ depend strongly upon temperature. The dependence of M$_{0}$ with T/T$%
_{C}$ evidence the FM transition at T/T$_{C}$ $\simeq $\ 1. As observed on
the temperature dependence of M$_{ZFC}$, the FM component first reaches a
maximum at T/T$_{C}$ $\simeq $ 0.8 that corresponds to T$_{a}$ (M$_{ZFC}$
cusp temperature) and decreases further with decreasing the temperature. The
FM component M$_{0}$ is larger for {\it x} = 0.10 compared to {\it x} =
0.15, in agreement with its stronger FM behavior. The variation of M$_{r}$
with T/T$_{C}$ shows that for {\it x} = 0.15 the relaxing component first
increases from T/T$_{C}$ $\simeq $\ 1 to T/T$_{C}$ $\simeq $\ 0.45, as a
consequence either of an increase of the clusters number or of the growth of
the clusters size, and decreases for T/T$_{C}$ 
\mbox{$<$}%
0.45. The resulting maximum in M$_{r}$ coincides with the temperature of the
frequency dependent lower maximum in the out-of-phase susceptibility, which
supports the idea that it is related to the beginning of cluster freezing. A
similar behavior is observed for {\it x} = 0.10, but at a lower temperature,
with the maximum in M$_{r}$ occurring around T/T$_{C}$ $\simeq $\ 0.20. The
relaxing component M$_{r}$ in {\it x} = 0.10 sample is also much lower than
in {\it x} = 0.15, due to a less pronounced glassy behavior. The relaxation
time $\tau _{r}$ is extremely sensitive to any noise in the data. However,
our fitting (not shown) indicate that $\tau _{r}$ increases with decreasing
temperature for both {\it x} = 0.15 and {\it x} = 0.10 , suggesting a
regular stiffening of the spin relaxation, which is compatible with an
increase of the clusters size and enhanced intercluster frustration.

To further investigate the dynamic magnetic behavior of these metallic glass
compounds the wait time {\it t}$_{w}$ dependence of the long time relaxed
magnetization, or aging effect, was studied. The results were obtained with
a small dc field, H = 0.3 mT. Additional measurements (not shown) as a
function of the applied field revealed that the magnitude of aging effect is
reduced when the field increased, and totally disappears at H = 10 mT.
Magnetization vs. time was measured with wait times {\it t}$_{w}$ = 100,
1000, and 10000 s, before application of the magnetic field. Results for the 
{\it x} = 0.10 sample, measured at T/T$_{C}$ = 0.2, are plotted in Fig. 7a.
This temperature was chosen because it corresponds to the maximum in the
relaxing component M$_{r}$. It is clear from the figure that the measured
magnetization strongly depends on {\it t}$_{w}$, which confirms the
existence of aging processes. Macroscopically, aging means that the system
becomes ``stiffer'' for larger wait time, i.e., the measured magnetization
is lower for higher {\it t}$_{w}$, as observed in the data. The
corresponding time dependent relaxation rate, S({\it t}) = $\partial $M/$%
\partial $(ln{\it t}), is plotted in Fig. 7b. The S(t) curves were obtained
by taking the derivative of a polynomial fit of the magnetization data. It
is readily observed that S({\it t}) reaches a maximum, corresponding to an
inflection point on the M({\it t})/M(0) curves, that shifts to longer
observation times for longer values of {\it t}$_{w}$. These results are
similar to those reported for various glassy systems. \cite
{Nan,NorbNdSr,CuMn,Jonas} It may be pointed out that the maximum in S({\it t}%
) is not centered on {\it t}$_{w}$ as expected. However, it is believed that
this is related to the additional time required for stabilizing the
temperature and the magnetic field applied.

For La$_{0.55}$Y$_{0.15}$Ca$_{0.3}$MnO$_{3}$, S({\it t}) has been calculated
at various temperatures, and the results are shown in Fig. 8. Aging effects
are observed at T/T$_{C}$ = 0.2, 0.4 and 0.6. The wait time dependence of
the peak in S({\it t}) clearly increases with temperature. Aging effects are
more pronounced at higher temperatures, i.e., when the FM state is stronger,
as reported previously for La$_{0.5}$Sr$_{0.5}$CoO$_{3}$\cite{Nan} or for
the two-dimensional ferromagnet Rb$_{2}$Cu$_{0.89}$Co$_{0.11}$F$_{4}$. \cite
{Schin} This may be explained by a more steady cluster growth in the FM
phase. Larger clusters give rise to a slower response due to larger
free-energy barriers and, correspondingly, a larger number of spins to be
simultaneously flipped; the smaller the cluster, the more rapidly it will
relax. At T/T$_{C}$ = 0.2 on the contrary, the relaxation is slow (Fig. 6b)
and the aging effect small, because the regular cluster growth with
decreasing temperature lead eventually to a blocking of the clusters at T$%
_{f}$ and below. It may also be noted in the data that the maximum in S({\it %
t}) occurs at higher observation times for lower temperatures. This effect
is mostly experimental in origin at low {\it t}$_{w}$ (10$^{2}$ seconds),
because {\it t}$_{w}$ is of the same order of the time it takes to reach the
measuring temperature, which effectively increases the wait time at low
temperatures. However, at high {\it t}$_{w}$ (10$^{4}$ seconds), the wait
time is much higher than the time for temperature stabilization, and the
shift in the maximum of S({\it t}) with temperature may be associated with
the slower response time of the clusters at low temperatures. It is worth
mentioning that the measuring field of 0.3 mT may not be in the linear
regime, which could affect these results. However, larger fields, outside
the linear regime, mostly change the intensity but not the position of the
maximum in S({\it t}).\cite{NorbNdSr}

\section{Conclusions}

The peak observed in the temperature dependence of the relaxing component of
the magnetization, M$_{r}$, or equivalently a maximum in the slope of M({\it %
t})/M(0) at a given temperature (Fig. 5b), combined with the aging effects
observed for La$_{0.55}$Y$_{0.15}$Ca$_{0.3}$MnO$_{3}$, and, to a lesser
extend for La$_{0.6}$Y$_{0.10}$Ca$_{0.3}$MnO$_{3}$, evidence the importance
of magnetic frustration in these samples and establish the existence of
cluster glass properties over a broad temperature range. The enhanced
lattice distortion resulting from the presence of cationic size mismatch on
the perovskite A-site weakens the Mn$^{3+}$-O-Mn$^{4+}$ FM double-exchange
interactions and favors frustration between the ferromagnetic and
antiferromagnetic superexchange interactions. This competition is sufficient
to suppress long range FM order, giving rise to the appearance of spatially
confined ferromagnetic clusters, which in turn are responsible for the
observed glassy behavior of the system due to frustrated interaction amongst
their magnetic moments. It is important to stress out that the observed
cluster glass behavior occurs within the metallic phase, as opposed to the
standard spin glass phase of the manganites, which occurs in the insulating
phase. From the behavior of M$_{r}$, it is inferred that clusters start to
form just below the FM transition T$_{C}$. As the temperature is lowered,
the size and/or number of these magnetic clusters increase, leading to a
maximum in the relaxing component of the magnetization. The slower (weaker)
relaxation at lower temperatures can be attributed to the freezing of the
clusters, similar to the blocking of the magnetic moments in a
superparamagnet, due to an enhanced intercluster frustration resulting from
oversized domains.

The overall relaxation results presented here demonstrate the existence of a
magnetic glassy behavior within the metallic phase, which plays an important
role in the physics of these compounds. An important issue which remains to
be verified is whether the observed cluster glass properties in La$_{0.7-x}$Y%
$_{x}$Ca$_{0.3}$MnO$_{3}$ develop uniformly distributed due to disorder or
enhanced canting within the FM phase, or possibly due to phase separated
spin glass regions in a FM background. It is also worth noting that M vs. H
measurements are similar to that obtained in RSG systems, indicating the
existence of more complex frustration effects and disorder in this system.
Within the phase segregation scenario, it is possible that the FM clusters
and the FM background give rise to different glassy behavior.

\section{Acknowledgments}

We thank F. Wolff and P. Pureur for assistance with the SQUID measurements,
and M.A. Gusm\~{a}o for helpful discussions. This work was partially
financed under the contract PRONEX/FINEP/CNPq no 41.96.0907.00. Additional
support was given by FUJB. The sample preparation was funded by the UK-EPSRC.

* Corresponding author; e-mail: freitas@if.ufrj.br

Figure 1 - Temperature dependence of the resistivity of the studied
compounds.

Figure 2 - Field-cooled (solid lines) and zero-field-cooled (dotted lines)
dc magnetization of La$_{0.7-x}$Y$_{x}$Ca$_{0.3}$MnO$_{3}$, measured with 5
mT.

Figure 3 - Temperature dependence of the coercive field, H$_{C}$, obtained
from M(H) measurements after the application of H = 9 T. The inset shows the
low field portion of the M(H) measurements taken at 2.0 K. The lines are
only guide to the eyes.

Figure 4 - Out-of-phase susceptibility ($\chi $'') of La$_{0.55}$Y$_{0.15}$Ca%
$_{0.3}$MnO$_{3}$, measured with an ac field h$_{ac}$ = 0.5 mT and
frequencies {\it f} = 90 (circles), 250 (squares), 700 (triangles), and 2000
Hz ( diamonds). The position of a frequency dependent feature in the data is
indicated by T$_{f}$. The inset shows the in-phase susceptibility ($\chi $')
of the same sample.

Figure 5 - (a) Normalized zero field cooled magnetization of La$_{0.7-x}$Y$%
_{x}$Ca$_{0.3}$MnO$_{3}$ measured as a function of time, with H = 5 mT at a
reduced temperature T/T$_{C}$ = 0.2; (b) The same measurement for La$_{0.55}$%
Y$_{0.15}$Ca$_{0.3}$MnO$_{3}$, at T/T$_{C}$ = 0.1, 0.4, and 0.8.

Figure 6 - Evolution of the fitting parameters M$_{0}$ and M$_{r}$ (see text
for details) as a function of a reduced temperature T/T$_{C}$, for {\it x} =
0.10 and 0.15 samples.

Figure 7 - (a) Zero field cooled relaxation magnetization, and (b) the
corresponding relaxation rate S({\it t}) = $\partial $M/$\partial $ln{\it t}%
, of La$_{0.60}$Y$_{0.10}$Ca$_{0.3}$MnO$_{3}$, measured at T/T$_{C}$ = 0.2
with H = 0.3 mT, after different wait times {\it t}$_{w}$ = 10$^{2}$, 10$%
^{3} $, and 10$^{4}$ s.

Figure 8 -Relaxation rate S({\it t}) = $\partial $M/$\partial $ln{\it t} of
La$_{0.55}$Y$_{0.15}$Ca$_{0.3}$MnO$_{3}$ measured after different wait times 
{\it t}$_{w}$ = 10$^{2}$ (open circles) and 10$^{4}$ (solid circles) at T/T$%
_{C}$ = 0.2, 0.4, 0.6. The arrows indicate the peak position of each data
curve.

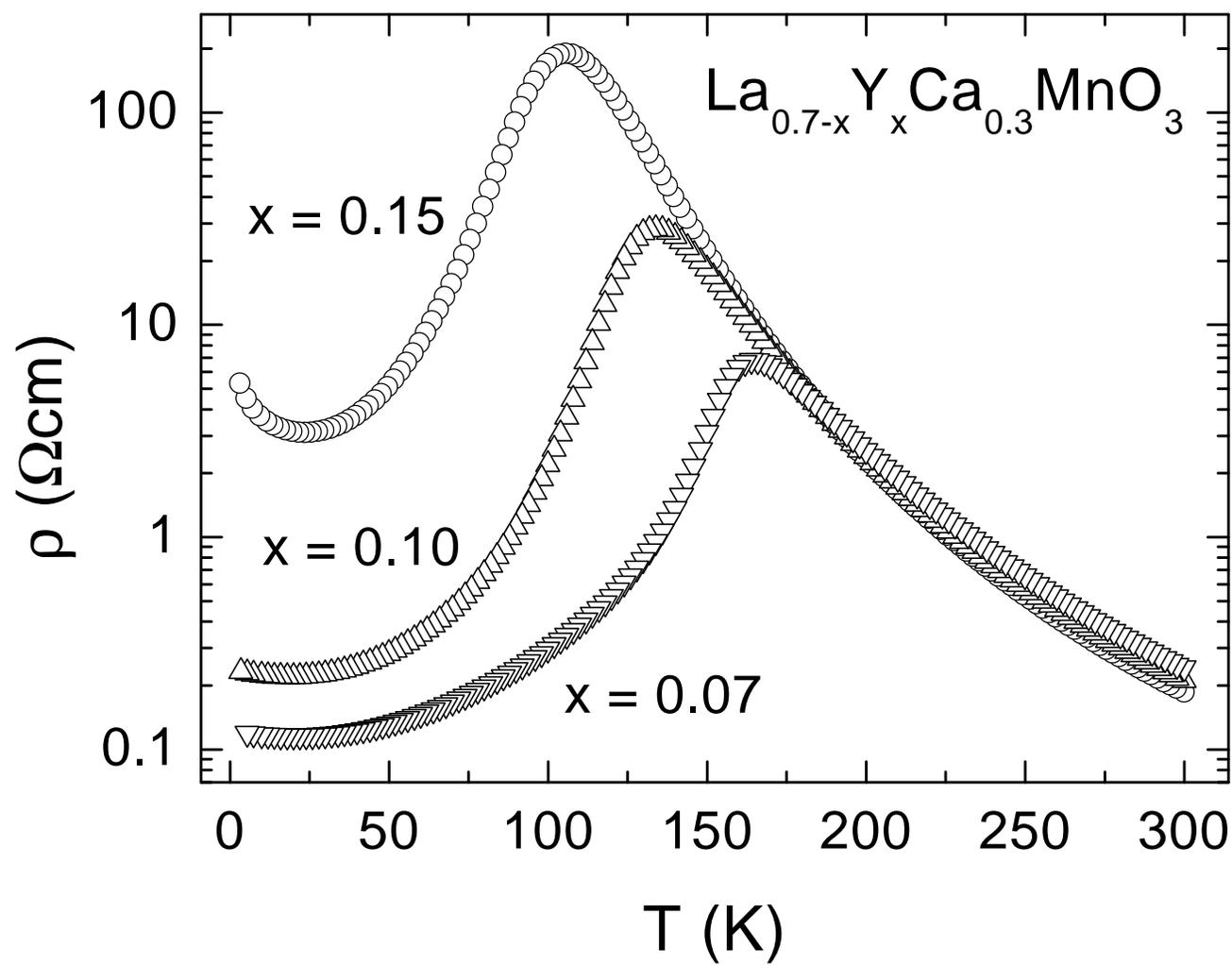
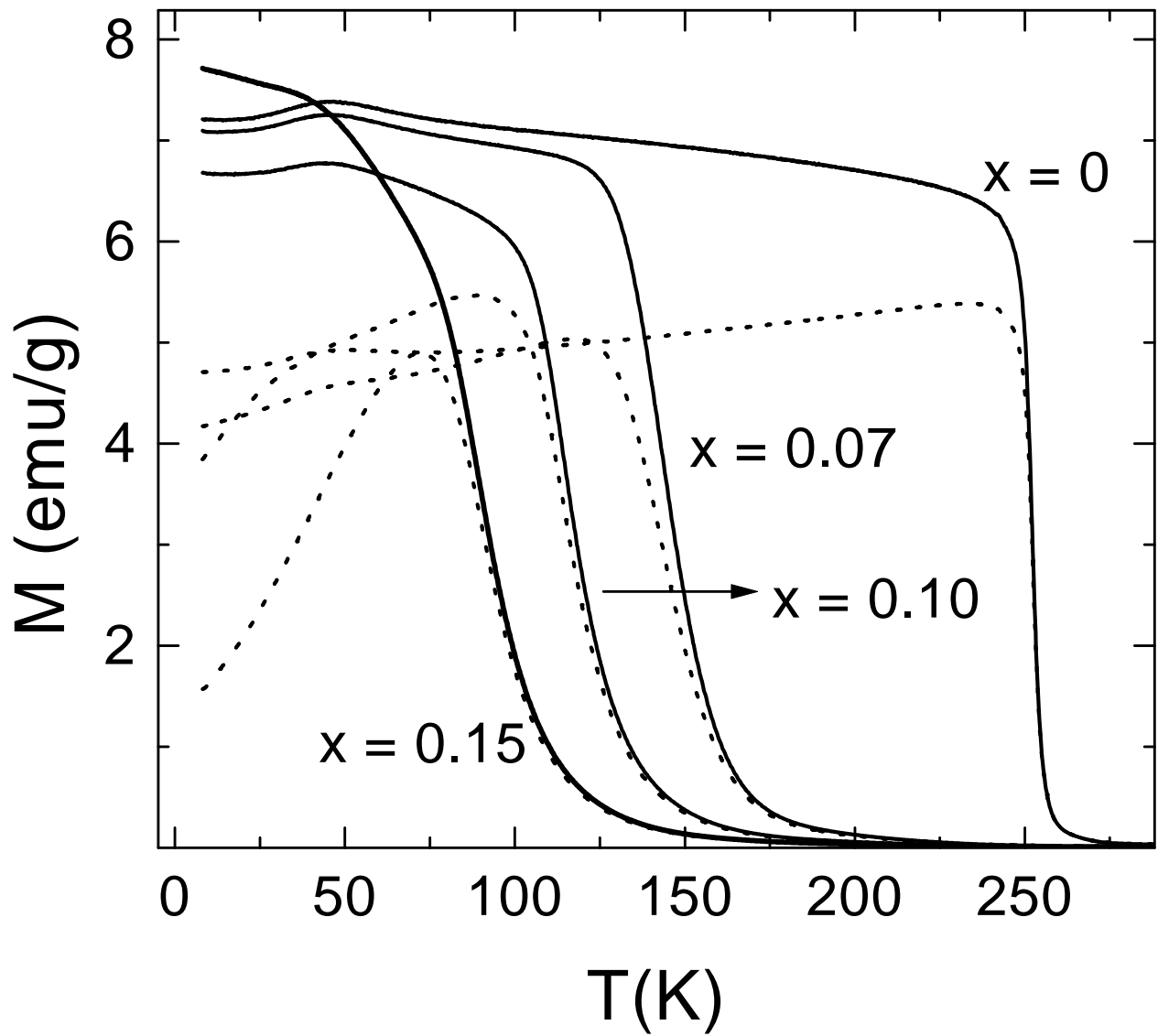
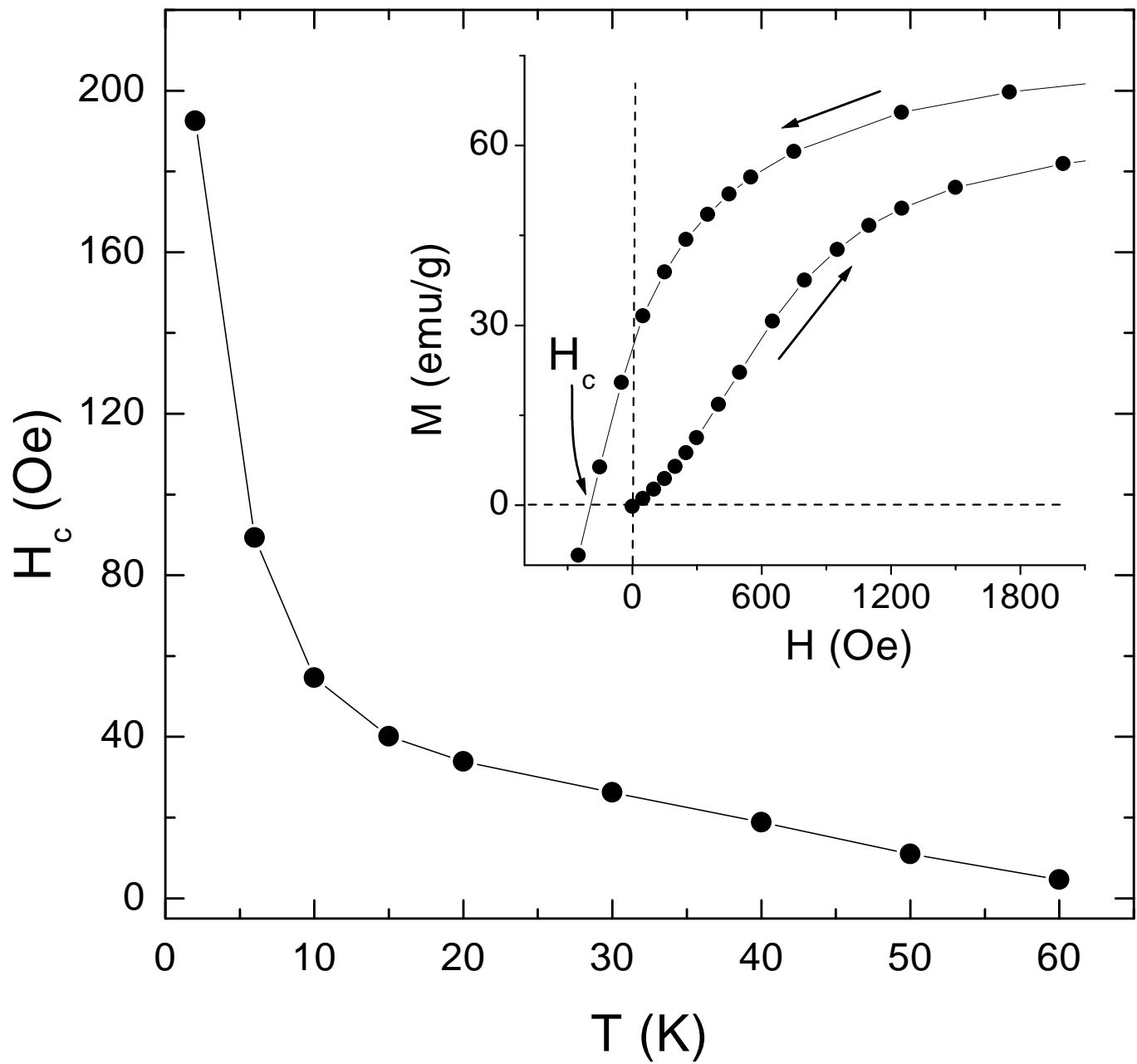
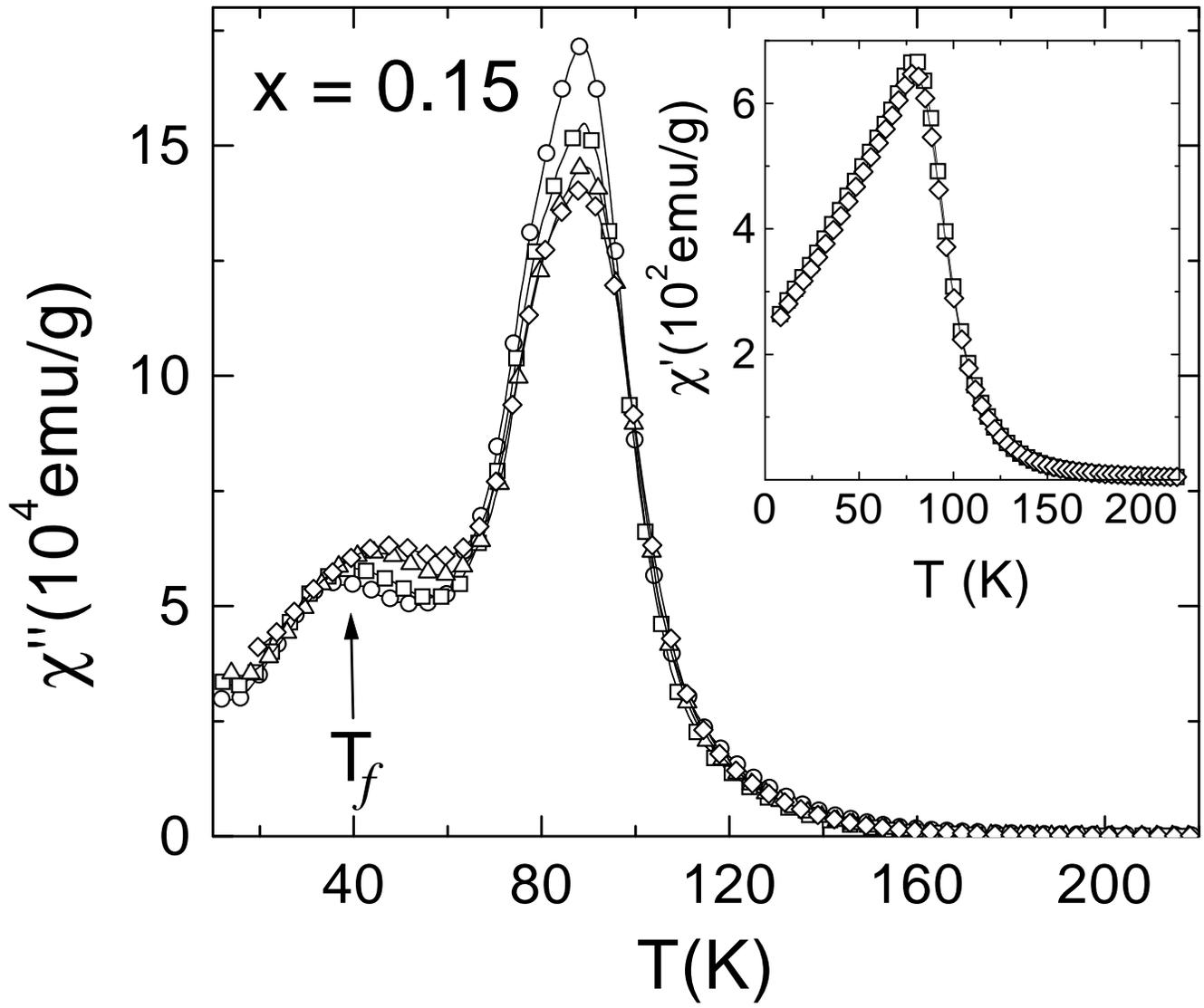
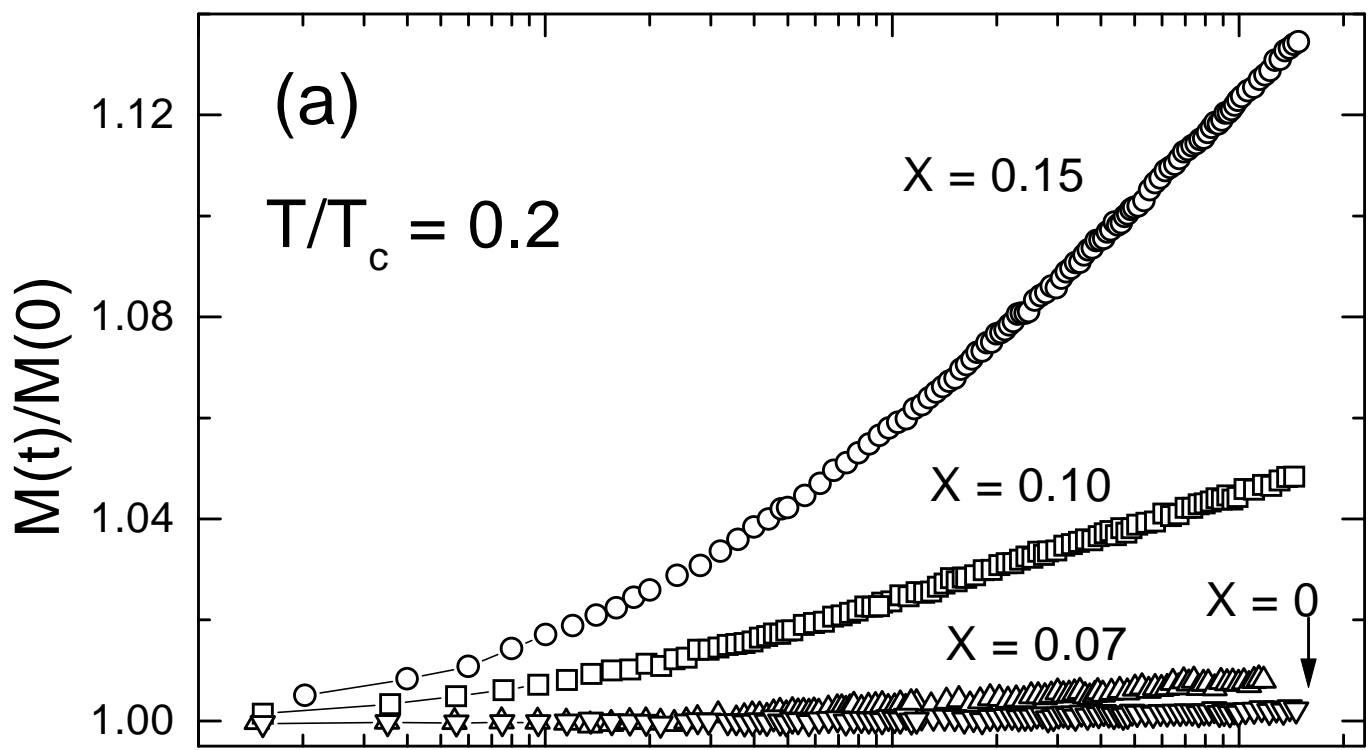
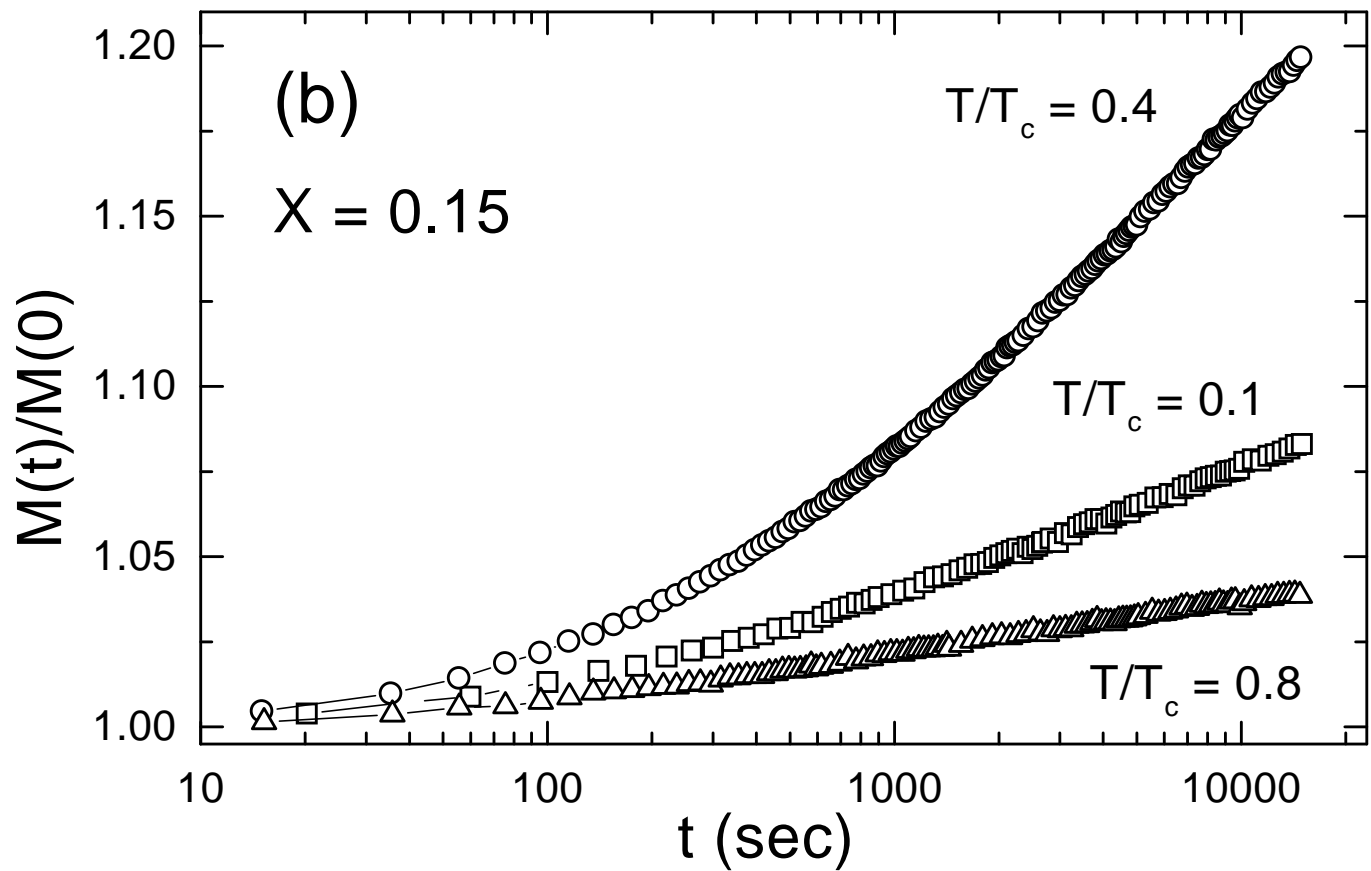
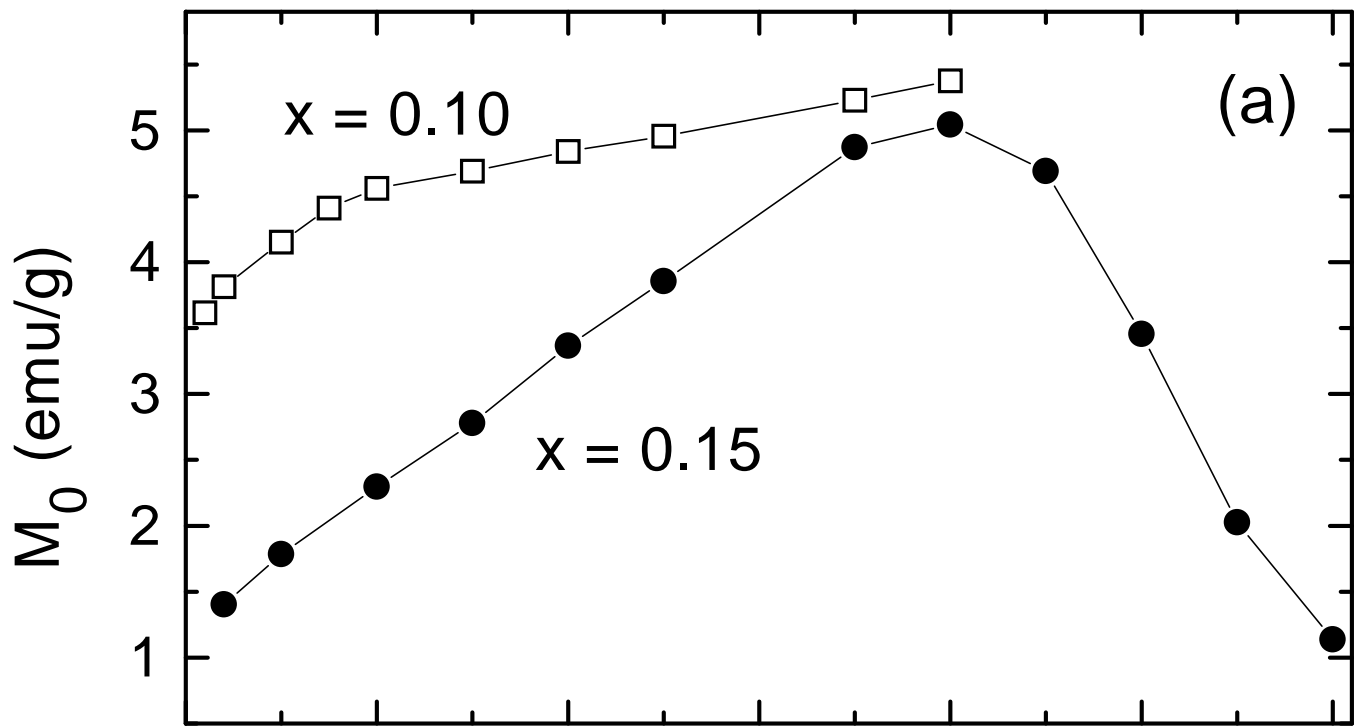
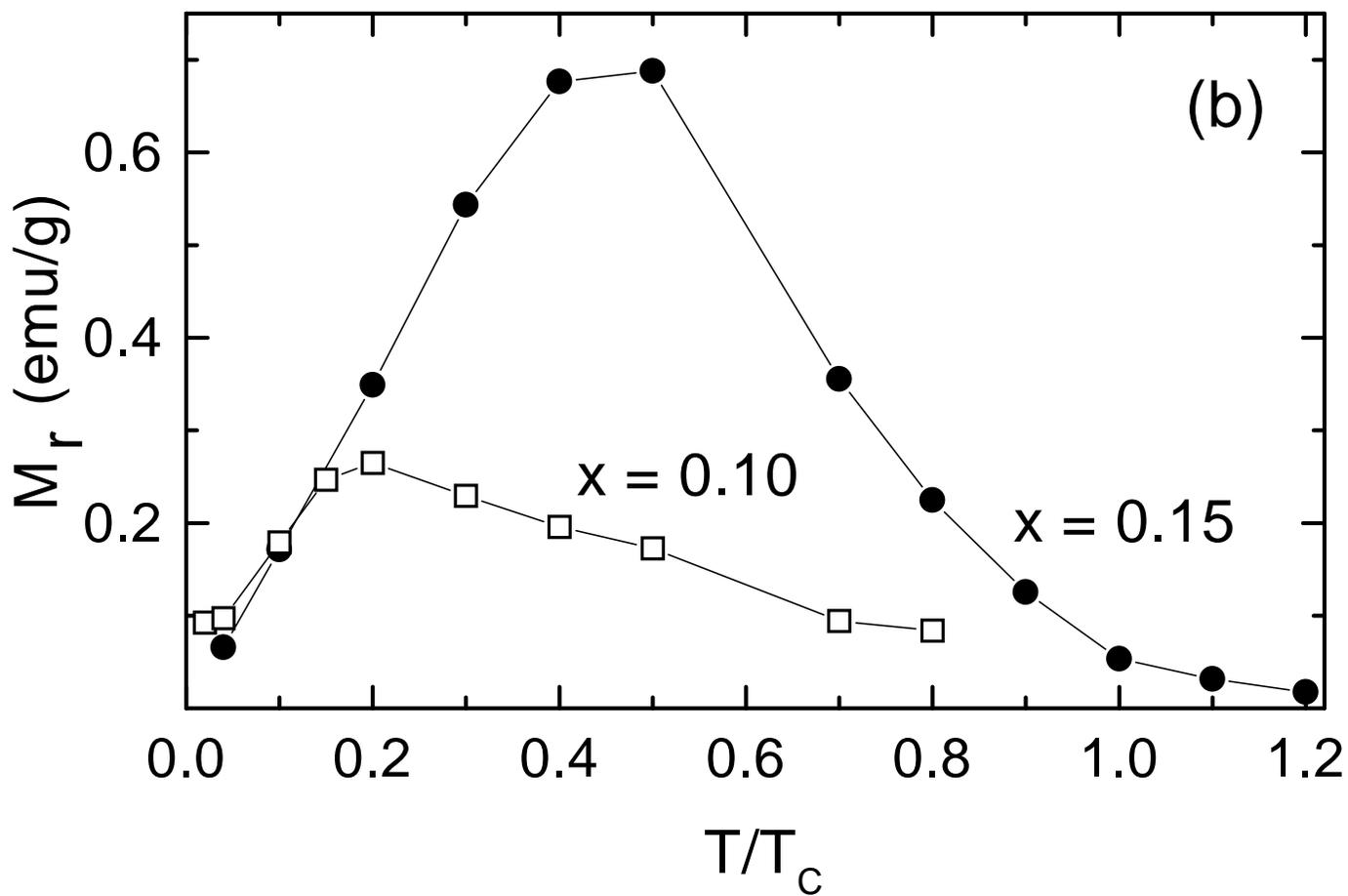
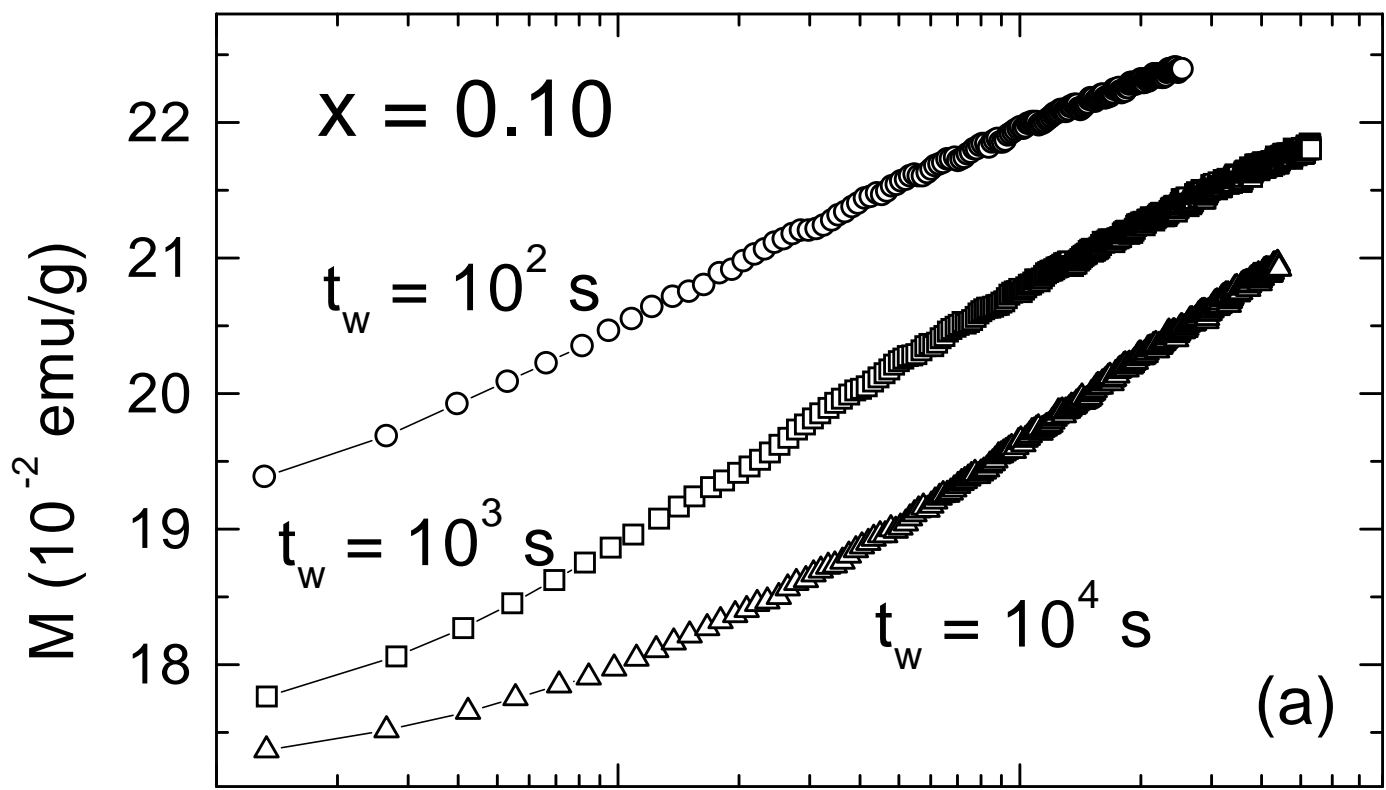
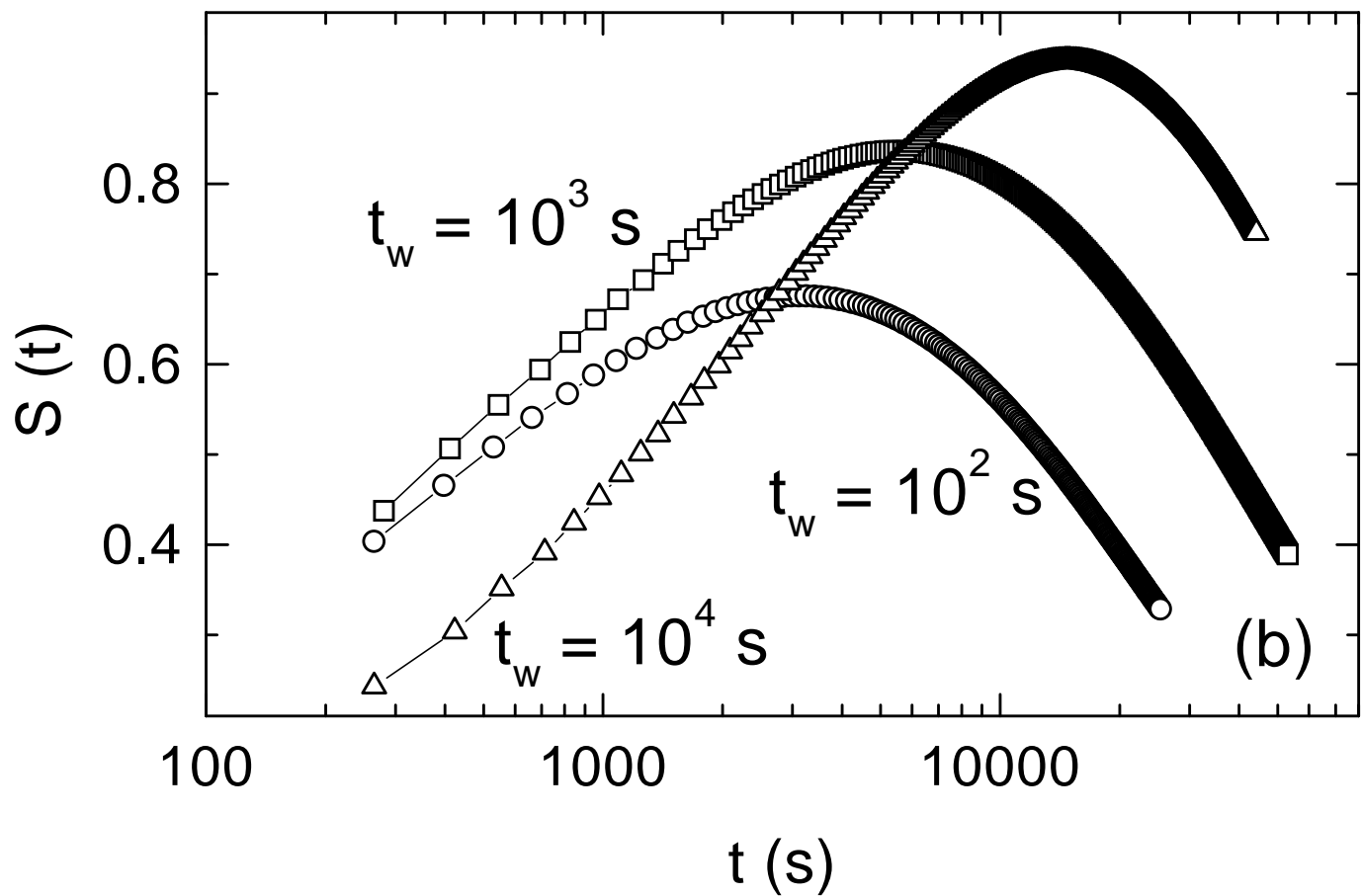
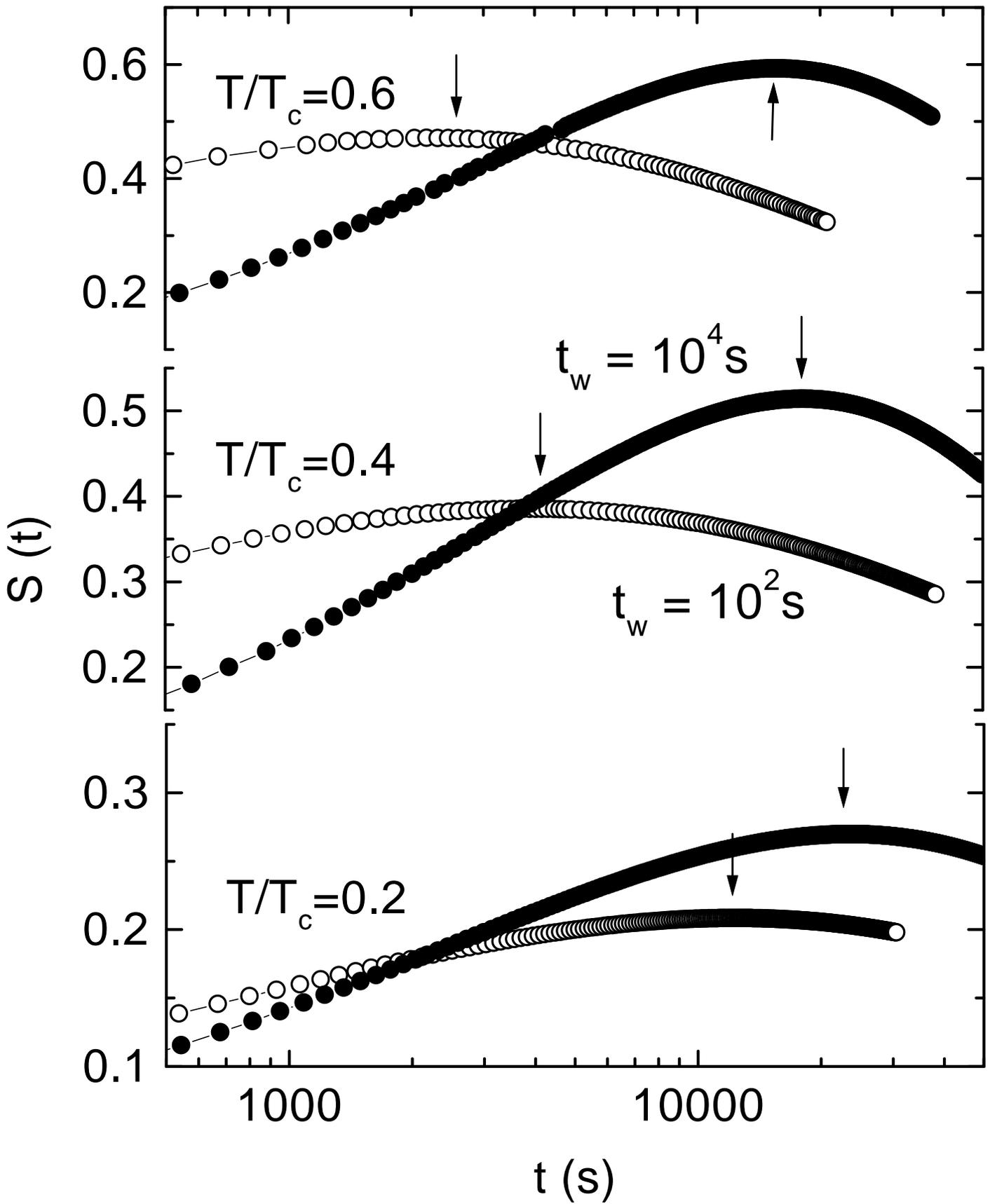
\end{document}